\documentclass{Interspeech2024}

\usepackage[table,xcdraw]{xcolor}
\usepackage{adjustbox}
\usepackage{multirow} 
\usepackage{graphicx}
\usepackage{booktabs}
\usepackage{hyperref}

\interspeechcameraready

\title{Prosody-Driven Privacy-Preserving Dementia Detection}

\name[affiliation={1}]{Dominika}{Woszczyk}
\name[affiliation={1,2}]{Ranya}{Aloufi}
\name[affiliation={1}]{Soteris}{Demetriou}

\address{
  $^1$Imperial College London, UK\\
  $^2$Taibah University, Saudi Arabia}
\email{d.woszczyk19@imperial.ac.uk, r.aloufi18@imperial.ac.uk, s.demetriou@imperial.ac.uk}


\begin{document}

\maketitle

\begin{abstract}
    

    Speaker embeddings extracted from voice recordings have been proven valuable for dementia detection. However, by their nature, these embeddings contain identifiable information which raises privacy concerns. In this work, we aim to anonymize embeddings while preserving the diagnostic utility for dementia detection. Previous studies rely on adversarial learning and models trained on the target attribute and struggle in limited-resource settings. We propose a novel approach that leverages domain knowledge to disentangle prosody features relevant to dementia from speaker embeddings without relying on a dementia classifier. Our experiments show the effectiveness of our approach in preserving speaker privacy (speaker recognition F1-score .01\%) while maintaining high dementia detection score F1-score of 74\% on the ADReSS dataset. Our results are also on par with a more constrained classifier-dependent system on ADReSSo (.01\% and .66\%), and have no impact on synthesized speech naturalness.\footnote{The code is available at ~\url{https://github.com/domiwk/privacy-preserving-ad-detection}}\footnote{Samples are available at~\url{https://shorturl.at/cNS39}}

\end{abstract}

\section{Introduction}
\label{sec:intro}

\vspace{0pt}\noindent\textbf{Problem Statement.} Advances in deep learning, combined with non-invasive biomarkers like speech, offer a promising opportunity for large-scale disease diagnosis. Researchers have investigated the use of speech signals for detecting different medical conditions, such as neurodegenerative diseases (e.g., Parkinson’s and Alzheimer’s)~\cite{de2020cross} and respiratory ailments (e.g., COVID-19)~\cite{ casanova2021transfer}.

Speaker embeddings (e.g., i-vector, x-vector, and ECAPA-TDNN) are a type of feature set utilized in the detection of early signs of diseases like dementia~\cite{botelho2022challenges}.
However, speaker embeddings often contain more information than required for their intended tasks, which poses potential privacy concerns. For example, 
they contain speaker-specific information, which makes them effective in automatic speaker verification (ASV) and zero-shot text-to-speech (TTS) among others.  This unintentional information leakage might violate GDPR policies~\cite{gdpr} and its data minimization principle~\cite{gdprSensitiveData} (i.e., ``adequate, relevant and limited to what is necessary in relation to the purposes for which they are processed''), leaving individuals vulnerable to discrimination, extortion, and targeted advertisements by third--parties. Moreover, since speaker embeddings can be used to predict both dementia and verify a speaker, they can be potentially classified as ``individually identifiable health information''. Under the HIPAA Privacy Rule~\cite{hipaa} such information is defined as ``protected health information'' (PHI) and must be properly de-identified.

\vspace{0pt}\noindent\textbf{Prior Works.} Anonymization or de-identification refers to the task of concealing the speaker’s identity while retaining the linguistic content, thereby making the data unlinkable~\cite{tomashenko2020introducing}. According to the ISO/IEC International Standard 24745 on biometric information protection~\cite{ISO/IEC}, biometric references must be irreversible and unlinkable for full privacy protection. Most of the proposed
works focus on protecting speaker identity, using
voice conversion (VC) mechanisms~\cite{ srivastava2020evaluating,noe2020adversarial}. 
Beyond speaker
identity, various works propose to protect speakers' attributes and paralinguistic information such as emotion~\cite{chouchane2023differentially,teixeira2023privacy}, gender ~\cite{noe2022bridge}, age~\cite{ teixeira2023privacy} or nationality~\cite{luu2022investigating}. Attribute obfuscation allows a speaker to conceal specific personal aspects in their voice representation while still maintaining overall performance~\cite{aloufi2020privacypreserving, noe2022bridge}.  
Adversarial training disentangles dimensions in latent spaces for speaker verification while minimizing detection of specific attributes~\cite{chouchane2023differentially}. 
The need of an external attribute classifier, especially for low-resourced attributes like dementia with no or limited training data, is a significant constraint. An alternative approach 
is to work at the feature level rather than the utterance level~\cite{noe2022bridge, lavania2023utility, tran23b_interspeech}. By extracting and sanitizing feature representations from speech, we can share privacy-aware features instead of complete utterances.
Noé et. al., in~\cite{noe2022bridge}, for example, proposed 
a Normalizing Flow-based architecture to disentangle sex information in x-vectors. 

\vspace{0pt}\noindent\textbf{Our Approach.} In this work, we leverage prosody disentanglement as a method for speaker anonymization in dementia detection. Specifically, our goal is to preserve the dementia attribute in speaker embeddings while reducing speaker-related information (identity). Often, adversarial training or multi-task learning is utilized to confine information within bottlenecks, separating out dimensions such as content, pitch, rhythm, and timbre~\cite{qian2020unsupervised,noe2020adversarial}. However, we take a different approach, focusing on prosodic features that are known to be prominent in dementia speech, such as articulation rate, pauses, and disfluencies. Our hypothesis is that by disentangling these features from speaker representations, we can effectively obscure the speaker's identity while minimizing any impact on dementia-related information. Our system achieves this without relying on dedicated classifiers, using domain knowledge and adversarial learning techniques. 

\vspace{3pt}\noindent Below we summarize the ~\textbf{main contributions} of this work:
\begin{enumerate}
\item \textbf{Novel Approach}: We propose a novel method for preserving privacy in speaker embeddings in low-resource settings through domain knowledge and prosody disentanglement.
\item \textbf{New Application Domain} We shed light on a sensitive attribute, dementia, that has not been extensively investigated in previous attribute obfuscation and anonymization works.
\end{enumerate}

\section{Privacy-Preserving Dementia Detection}
\label{sec:meth}

\subsection{Threat Model}

Our threat model focuses on medical speech-processing systems developed for detecting dementia and their handling.
These systems use audio data to generate embeddings that can help identify signs of dementia. This raises significant privacy concerns as speaker embeddings can potentially reveal individuals' identities. We consider an adversary that has access to the anonymized embeddings and aims to re-identify the user by using the embeddings not for their primary purpose (dementia classification) but for speaker recognition. This information could be exploited for discriminatory purposes or targeted advertising. Effective anonymization techniques should prevent such linkage attacks while preserving speech naturalness, intelligibility, and performance in dementia detection.



We aim to safeguard the privacy of user identity in different scenarios that involve voice embeddings or medical speech processing. This involves obfuscating any identifying information that a user may not want to share, without compromising the functionality of the system. We also emphasize the need to offer various privacy settings to balance the trade-off between privacy and utility, and to encourage transparent privacy management practices.

\subsection{Proposed Approach}

We devise a prosody-based privacy-preserving extraction for speaker representations, trained on a larger auxiliary dataset that does not need to have the target attribute label. We leverage domain knowledge for the task of dementia detection and propose a method that performs disentanglement that focuses on prosody features relevant to dementia. Indeed, previous studies have shown that features such as speech rate, mean energy levels, number of pauses and lengths~\cite{vincze2021telltale, pastoriza2022speech} are informative for dementia classification. We explore two approaches: adversarial learning and mutual information-guided shuffling.
\subsection{Adversarial Learning for Speaker-Prosody Disentanglement}

The proposed adversarial model is based on domain adversarial training~\cite{ganin2016domain} which aims to create domain-invariant latent spaces by maximising the domain discriminator's confusion while minimizing the task-specific loss. In our case, we train our model to extract dementia-relevant prosody features while maximising the loss of speaker-relevant features. To train this model, we can take advantage of a larger dataset and extract prosodic features. At inference time, we run the model on any dataset we wish to anonymize. The model consists of several components: the feature extractor that extracts speaker representations, and prosody regressors for each prosodic feature. The loss of our model is defined as:

\begin{equation}
\mathcal{L} = \sum_{i=0}^{\text{n}}\mathcal{L}_{\text{ADpros}_{i}} - \sum_{j=0}^{\text{n}}\lambda_{j}\mathcal{L}_{\text{SPKpros}_{j}}
\end{equation}

where $\mathcal{L}_{\text{ADpros}_{i}}$ denotes the $i^{th}$ loss associated with the $i^{th}$ dementia-relevant prosody feature extraction, and $\mathcal{L}_{\text{SPKpros}_{j}}$ the $j^{th}$ loss associated with the $j^{th}$ speaker-relevant prosody feature regressor. $\lambda_{j}$ is a hyperparameter that controls the balance between the different objectives and $n$ is the total number of prosody regressors.



\subsection{Mutual Information-Guided Shuffling}

We propose a feature selection approach based on mutual information to identify important features relevant to dementia while perturbing less relevant features to lower speaker recognition accuracy. Our method is similar to~\cite{lavania2023utility} where they extract Shapley values from a classifier to select key dimensions, however, unlike theirs, our approach is \textit{classifier-free}. The intuition behind this approach is that critical features associated with dementia are likely to have higher mutual information with the target variable (dementia), while shuffling the remainder to preserve dementia-related features but decrease speaker recognition accuracy. We also explore using prosody features instead of the dementia label as the target variable.

Mutual information is a measure of mutual dependence between two random variables. In our context, we compute the mutual information between the distribution of dimension of the speaker embeddings and the distribution of the dementia label across the dataset. Formally, the mutual information \(I(X; Y)\) between random variables \(X\) and \(Y\) is defined as:

\begin{equation}
I(X; Y) = \sum_{x \in X} \sum_{y \in Y} p(x, y) \log \left( \frac{p(x, y)}{p(x)p(y)} \right)
\end{equation}

where \(p(x, y)\) is the joint probability mass function of \(X\) and \(Y\), and \(p(x)\) and \(p(y)\) are the marginal probability mass functions of \(X\) and \(Y\), respectively. To estimate the mutual information, we use the k-Nearest-Neighbour-based MI estimator~\cite{ross2014mutual} which can be applied to both continuous and discrete variables. We design the feature selection strategy as follows: 1) Compute the mutual information between each embedding dimension and the dementia/prosody variable across the whole corpus. 2) Select top $n$ dimensions as important dementia-related features. 3) Shuffle the remaining features. When combining several features, compute the top $n$ dimensions for each and compute the union as the set of important features.


\section{Experimental Setup}
\label{sec:exp}

\subsection{Datasets}
\label{sec:datasets}


We train our disentanglement model on the LibriSpeech dataset~\cite{panayotov2015librispeech}, a corpus of read English speech. We selected the \texttt{train-clean-100} subset which consists of 100 hours of speech from 251 speakers. We split the data into 25685, and 2850 samples for train, and dev sets respectively. We extract prosody features for each sample. 

For testing dementia detection, we use two publicly available datasets widely studied in dementia classification: ADReSS~\cite{luz2020alzheimer} and ADReSSo~\cite{luz2020alzheimer}. The ADReSS (ADR) dataset is a subset of the DementiaBank dataset, a collection of recordings from control (CD) and dementia (AD) patients describing the Cookie Theft Picture~\cite{becker1994natural}. 
Manual transcripts are provided and we split each sample into segments using the sentence-level timestamps from the transcripts. We group the original test and train sets and split the data to contain one sample per speaker in the test set and keep the remaining samples as training data. We end up with 1723 samples (868 CC $|$ 855 AD) in the training set and 156 (78 CC $|$ 78 AD) in the test and validation sets. The ADReSSo (ADRo) dataset is another subset of DementiaBank designed for detecting dementia from spontaneous speech only, without access to manual transcriptions. The original set consists of 151 train samples (87 CC $|$ 74 AD) and 71 test samples (35 CC $|$ 35 AD). We use the provided segmentation timestamps to isolate segments spoken by the patients and get 2705 samples in the train set (920 CC $|$ 1026 AD) and 231 samples (74 CC $|$ 87 AD) in the test and validation sets, one per speaker.   
\subsubsection{Data Processing}
 For the dementia datasets, we split samples into segments as described in Section~\ref{sec:datasets}. For both LibriSpeech and Dementia datasets, we extract a series of articulation features, number and length of pauses, f0 and mean energy. We compute the features with Parselmouth and Praat scripts made available by Feinberg~\cite{Feinberg_2022} and normalise them to be within [0,1]. When extracting embeddings, we trim the segments to 30s.

\subsection{Implementation Details}
\label{sec:implt}
We select a pre-trained ECAPA-TDNN~\cite{desplanques2020ecapa} embedding extractor, regarded as the state-of-the-art (SOTA) for speaker embeddings. We use the SpeechBrain~\cite{speechbrain} implementation and pretrained model which was trained on the VoxCeleb dataset. We augment the ECAPA-TDNN embedding extractor with classifiers for each prosodic feature, each consisting of two layers with 126 hidden dimensions. We use the Mean-Square-Error Loss for the prosody regressors and investigate $\lambda$ values in \{1,5,10,30\} and set $\lambda$ to 1.0. We train the model with a batch size of 8 with cyclical learning rate policy (CLR) and set the base and maximum learning rates to 1$\text{e-}7$ and 1$\text{e-}5$ respectively. We finetune the models for 15 epochs with early stopping. We compare our model to several models: a model trained on a dementia dataset to classify dementia while fooling the speaker classifier (ADV SPK$_{AD}$ + AD),  a model trained on the LibriSpeech dataset to fool the speaker classifier (ADV SPK$_{LS}$), a random shuffling method (Shuffle$_{Random}$) and a shuffling from~\cite{lavania2023utility} where we train an XGBoost model~\cite{Chen:2016:XST:2939672.2939785} on dementia and select top $n$ Shapley values (Shuffle$_{Shap}$). For the mutual information shuffling method (Shuffle$_{MI}$), we compute the mutual information using the sklearn library ~\cite{pedregosa2011scikit}. The dementia classifier is a two-layer model with a hidden space size of 8 and a ReLu activation between layers, trained with Binary Cross-Entropy. The speaker classifier is a two-layer model with a hidden space size of 96 trained with Cross-Entropy Loss. We optimise regressors with bayesian search.

\subsection{Evaluation Metrics}
\vspace{3pt}\noindent\textbf{Privacy.} We measure the privacy gain through the drop of an adversary's (speaker classifier from Section~\ref{sec:implt}) F1-score and speaker verification Equal Error Rate (EER) score (using cosine distance). Due to the size of our dataset, we focus on a black-box setting and do not adapt our adversary. We evaluate the anonymized embeddings against an SVM with a radial basis function kernel, as it performed the best on both datasets.


\vspace{3pt}\noindent\textbf{Utility.} 
To evaluate the ability to detect dementia, we train a neural dementia classifier on top of the embeddings and report the F1-score. We implement a feed-forward network with two layers, a hidden space size of 8 and a ReLu activation between layers. 
Additionally, we perform zero-shot speech generation with the new embeddings and measure MOSNet [41], SI-SDR [42], and STOI [43], which are used as objective metrics to evaluate the quality of speech signals, and WER (Word Error Rate) with Whisper~\cite{radford2023robust}. These metrics are used to assess the quality of the processed signal compared to the original signal. We pick YourTTS~\cite{casanova2022yourtts} and SpeechT5~\cite{ao-etal-2022-speecht5}, two open-source zero-shot TTS systems that have shown SOTA performance. 


\section{Results}
\label{sec:results}

\subsection{Prosodic Features Selection}
\begin{figure}[!h]
\centering
\includegraphics[width=0.9\columnwidth]{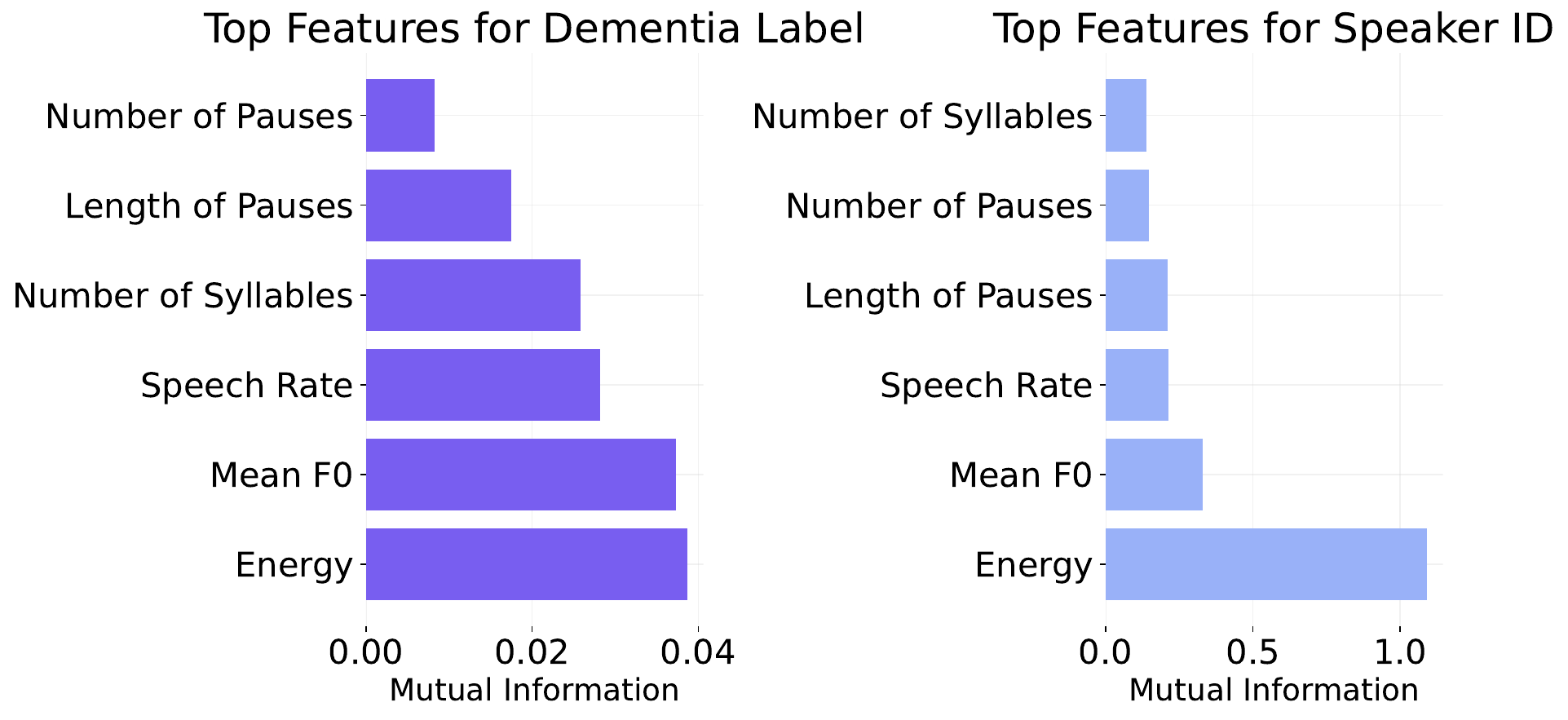}
\caption{Mutual information scores of key prosodic features extracted from audio segments for dementia label (left) and speaker identity (ID) (right) on the ADReSS dataset.}
\label{fig:mi_feats}
\vspace{-5pt}
\end{figure}

We choose prosody features based on mutual information (MI) with dementia labels in the ADReSS dataset, selecting the top 35 quantiles. Then, we assess their importance in speaker recognition by computing MI with speaker labels. Figure~\ref{fig:mi_feats} shows MI scores for features w.r.t. class and speaker identity. Notably, mean f0 and energy are significant for both dementia and speakers. However, mean energy is strongly linked with speaker identity, while f0 is a known speaker characteristic. To remove speaker information, we decide to separate features into prosodic and disfluency features to disentangle speakers and dementia. We select speech rate (spr), number of pauses (pnum), length of pauses (plength) and number of syllables as dementia features and mean f0 (f0) and energy (nrg) as speaker-relevant for the ADV PROS and Shuffle$_{MI\_pros}$ systems.

\subsection{Overall Results}

\begin{table}[htbp]
    \centering
    \Large
    \caption{Comparison of anonymization systems' performance. We report the dementia detection F1-score (AD), speaker recognition F1-score (SPK), their 95\% confidence intervals and Equal Error Rate (EER).}
    \label{tab:overall_res}
    \resizebox{\columnwidth}{!}{%
    \begin{tabular}{l|ccc|ccc}
        \toprule
        \multirow{2}{*}{\textbf{System}} & \multicolumn{3}{c|}{\textbf{ADReSS}} & \multicolumn{3}{c}{\textbf{ADReSSo}} \\
        \cmidrule(lr){2-4} \cmidrule(lr){5-7}
         & \textbf{AD $\uparrow$} & \textbf{SPK$\downarrow$} & \textbf{EER (\%)$\uparrow$} & \textbf{AD $\uparrow$ } & \textbf{SPK$\downarrow$} & \textbf{EER (\%)} $\uparrow$ \\
        \midrule
        Original & .81\large{ $\pm$ .06} & .33\large{ $\pm$ .06} & 35 & .73\large{ $\pm$ .06}& .36\large{ $\pm$ .04}& 35\\
        \midrule
        ADV SPK$_{AD}$ + AD & .64\large{ $\pm$ .06} & .00\large{ $\pm$ .02} & 47 & .63\large{ $\pm$ .06}& .00\large{ $\pm$ .02}& 45\\
        ADV SPK$_{LS}$ & .80\large{ $\pm$ .13}& .69\large{ $\pm$ .01}& 15 & .75\large{ $\pm$ .1}& .75\large{ $\pm$ .01}& 16\\
        ADV$_{nrg}$ PROS& .73\large{ $\pm$ .13}& .01\large{ $\pm$ .01}& 43 & .66\large{ $\pm$ .10}& .01\large{ $\pm$ .01}& 44\\
        \midrule
        Shuffle$_{Random}$ & .63\large{ $\pm$ .08}& .01\large{ $\pm$ .00}& 41 & .61\large{ $\pm$ .06}& .02\large{ $\pm$ .04}& 41\\
        Shuffle$_{Shap}$ & .62\large{ $\pm$ .07}& .02\large{ $\pm$ .02}& 41 & .64\large{ $\pm$ .07}& .04\large{ $\pm$ .02}& 40 \\
        Shuffle$_{MI\_AD}$ & .63\large{ $\pm$ .07}& .02\large{ $\pm$ .02}& 41 & .62\large{ $\pm$ .06}& .02\large{ $\pm$ .01}& 42 \\
        Shuffle$_{MI\_pros}$ & .62\large{ $\pm$ .08}& .01\large{ $\pm$ .02}& 44 & .68\large{ $\pm$ .06}& .02\large{ $\pm$ .01}& 43 \\
        \bottomrule
    \end{tabular}%
    }
\end{table}

Table~\ref{tab:overall_res} shows the performance of the anonymization systems. We see that ADV SPK$_{AD}$ + AD drops the speaker F1-score to 0\% for both ADR and ADRo and increases the EER by 12\%, 10\% respectively, giving the best anonymization performance. We note that the EER on the original samples is quite high (35\%), which we attribute to the noisy nature of the dataset. However, in this work, we are interested in the relative increase. We see the system reduces the AD detection F1-score to 65\%, on par with the shuffling approaches. Through experimentations, we observe that the energy (nrg) gave us stronger privacy guarantees across subsets, as shown in Table~\ref{tab:ablation_res}. We report ADV$_{nrg}$ PROS as our best system. ADV$_{nrg}$ PROS, drops the accuracy to 0.01\% and adds 8\% to the EER while preserving 73\% F1-score on ADR, a drop of 7\% from the original performance, making it the best-performing approach. On ADRo it achieves comparable performance to ADV SPK$_{AD}$ + AD and shuffling approaches. We compare it to the model trained adversarially against a speaker classifier ADV SPK$_{LS}$. Surprisingly, the models improved the speaker recognition and dementia classifiers on both datasets. We evaluated it in combination with other features, and a similar pattern was seen, hence we report only ADV SPK$_{LS}$. Finally, we evaluate different shuffling approaches. We report the systems selecting top 50 features as it gave us the best privacy/utility tradeoff for all systems. We note that the selection process with MI based on the label (Shuffle$_{MI\_AD}$) and prosody (Shuffle$_{MI\_pros}$) seems to add little to the accuracy when compared to Shuffle$_{Random}$. Nevertheless, the shuffling approaches achieve similar results to the strong baseline ADV SPK$_{AD}$ + AD,  while being simpler.

\subsection{Zero-shot TTS-based Evaluation}
The Voice Privacy Challenge (VPC) revealed that all methods, including x-vector embeddings and signal processing techniques, could lead to a degradation in speech naturalness and intelligibility~\cite{tomashenko2020introducing, meyer22b_interspeech}. 
We thus conducted an objective evaluation on the ADReSS dataset using speech synthesis models using two text-to-speech (TTS) systems: SpeechT5 and YourTTS (zero-shot TTS systems) to assess the effectiveness of anonymized embeddings in text-to-speech (TTS) tasks. We utilize sentence-level transcriptions as input data and condition the TTS system on speaker embeddings. For a fair comparison, we regenerate the raw recordings using the raw embedding extracted from the original recordings. We then use the anonymized embeddings of different settings to generate different test sets. 
Rankings were based on average performance across all metrics, with lower average ranks indicating better overall speech quality. MOSNet represents the Mean Opinion Score predicted by a neural network model for the quality of speech signals. Higher MOSNet, SI-SDR, and STOI scores indicate better quality and speech intelligibility. Table~\ref{tab:tts_results} reports the results of our proposed method under different settings and the baseline (synthesized raw recordings) using two TTS systems. We find that the anonymization results in
almost the same voice distinctiveness as the data originally had, and, according to speech recognition, produces intelligible speech recordings. ADV SPK$_{AD}$ + AD seems to strike the best balance in maintaining voice distinctiveness while ensuring anonymity, as shown by its lower Avg. Rank, even though all methods introduce some level of degradation in SI-SDR and WER, which is a trade-off for achieving speaker anonymity.

\begin{table}[]
\caption{Zero-Shot TTS Quality Metrics and WER on the ADReSS dataset.}
\label{tab:tts_results}
\Large
\begin{adjustbox}{width=\columnwidth}
\begin{tabular}{ccccccc}
\toprule
\textbf{TTS} & \textbf{{System}} & \textbf{{MOSNet$\uparrow$}} & { \textbf{SI-SDR$\uparrow$}} & { \textbf{STOI$\uparrow$}} & { \textbf{Avg. Rank$\downarrow$}} & { \textbf{WER (\%)$\downarrow$}} \\ 
\midrule
\multirow[c]{5}{*}{\textbf{SpeechT5}} & {Original} & 2.84 & -69.01 & .26 & 3.33 & 23 \\ \cmidrule(lr){2-7}
&{ ADV SPK$_{AD}$ + AD} & 2.84 & -73.41 & .23 & 1.83 & 30 \\
&{ ADV$_{nrg}$ PROS} & 2.84 & -72.23 & .21 & 2.83 & 35 \\
&{ Shuffle$_{MI\_pros}$} & 2.84 & -70.73 & .26 & 2.50 & 21 \\
&{Shuffle$_{Shap}$} & 2.84 & -72.16 & .23 & 2.83 & 34 \\
\midrule
\multirow[c]{5}{*}{\textbf{YourTTS}} & { Original} & 2.77 & -68.57 & .24 & 2.33 & 27 \\ 
\cmidrule(lr){2-7}
&{ ADV SPK$_{AD}$ + AD} & 2.82 & -68.60 & .23 & 1.33 & 30 \\
&{ ADV$_{nrg}$ PROS} & 2.82 & -67.90 & .23 & 2.83 & 29 \\
&{ Shuffle$_{MI\_pros}$} & 2.82 & -68.43 & .23 & 2.00 & 27 \\
&{ Shuffle$_{Shap}$} & 2.82 & -68.23 & .23 & 2.66 & 28 \\ \bottomrule
\end{tabular}
\end{adjustbox}
\end{table}

\subsection{Ablation study}

We investigate the impact of dementia-relevant and speaker-relevant prosodic features and report the results in Table~\ref{tab:ablation_res}. We find that the addition of prosody features improves dementia detection with a negligible impact on privacy metrics. Nearly all features are useful for preserving dementia, while the speaking rate did not seem to affect the detection. 

\begin{table}[htbp]
    \centering
    \Large 
    \caption{Different prosody sets. We report the dementia detection F1-score (AD), speaker recognition F1-score (SPK), their 95\% confidence intervals, and Equal Error Rate (EER).}
    \label{tab:ablation_res}
   \resizebox{\columnwidth}{!}{%
   \begin{tabular}{l|ccc|ccc}
        \toprule
        \textbf{System} & \multicolumn{3}{c|}{\textbf{ADReSS}} & \multicolumn{3}{c}{\textbf{ADReSSo}} \\
        \cmidrule(lr){2-4} \cmidrule(lr){5-7}
         & \textbf{AD$\uparrow$} & \textbf{SPK$\downarrow$} & \textbf{EER (\%)$\uparrow$} & \textbf{AD$\uparrow$} & \textbf{SPK$\downarrow$} & \textbf{EER (\%)$\uparrow$} \\
        \midrule
                ADV$_{f0}$ & .79\large{ $\pm$ .13}& .17\large{ $\pm$ .02}& 29 & .72\large{ $\pm$ .11}& .20\large{ $\pm$ .01}& 29\\
        ADV$_{f0}$ PROS & .76\large{ $\pm$ .13}& .29\large{ $\pm$ .02}& 29& .74\large{ $\pm$ .11}& .27\large{ $\pm$ .02}& 29\\
        ADV$_{nrg\_f0}$ & .73\large{ $\pm$ .13}& .03\large{ $\pm$ .02}& 32& .72\large{ $\pm$ .11}& .03\large{ $\pm$ .01}& 33\\
        ADV$_{nrg\_f0}$ PROS & .74\large{ $\pm$ .13}& .02\large{ $\pm$ .02}& 32& .68\large{ $\pm$ .11}& .03\large{ $\pm$ .01}& 33\\
        \hline
        ADV$_{nrg}$ & .65\large{ $\pm$ .11}& .01\large{ $\pm$ .01}& 45& .67\large{ $\pm$ .11}& .01\large{ $\pm$ .01}& 46\\
        ADV$_{nrg}$ PROS & .73\large{ $\pm$ .13}& .01\large{ $\pm$ .01}& 43 & .66\large{ $\pm$ .11}& .01\large{ $\pm$ .01}& 44\\
        \hline
        ADV$_{nrg}$ PROS$_{no\_nsyll}$ & .68\large{ $\pm$ .13}& .01\large{ $\pm$ .01}& 43& .66\large{ $\pm$ .11}& .01\large{ $\pm$ .01}& 45\\
        ADV$_{nrg}$ PROS$_{no\_plength}$ & .70\large{ $\pm$ .13}& .01\large{ $\pm$ .01}& 44& .64\large{ $\pm$ .11}& .00\large{ $\pm$ .01}& 47\\
        ADV$_{nrg}$ PROS$_{no\_pnum}$ & .69\large{ $\pm$ .13}& .00\large{ $\pm$ .01}& 44& .61\large{ $\pm$ .11}& .01\large{ $\pm$ .01}& 46\\
        ADV$_{nrg}$ PROS$_{no\_spr}$ & .71\large{ $\pm$ .13}& .02\large{ $\pm$ .02}& 43& .64\large{ $\pm$ .11}& .01\large{ $\pm$ .01}& 44\\
        \bottomrule
    \end{tabular}%
    }
    \vspace{-5pt}
\end{table}


\section{Discussion}
\label{sec:discussion}
In this work, we show that we can successfully anonymize speaker embeddings and preserve dementia detection by disentangling prosodic features relevant to dementia. Albeit the original EER is quite high, its increase demonstrates the effectiveness of our approach. Future work will explore a stronger adversary in a white-box setting and more robust to challenging speech. Furthemore, the approach can be extended to other attributes or health conditions. Nevertheless, a limitation of our work is the domain knowledge and feature analysis required to fine-tune the disentanglement. For our use case, the mean energy was an important feature in removing speaker information. However, this is dataset-specific and the generalizability of our system would need to be evaluated on a larger dataset, and explore more features. In terms of feature selection on the embeddings, future work would investigate explainable methods to understand what embeddings encode throughout the extraction process to better extract/obfuscate information. Finally, dementia has an important impact on linguistics and their content which could be further explored to preserve dementia better. However, this also raises other privacy concerns where patient transcripts might be leaked and calls for content disentanglement as well. We leave the exploration of this problem to future work.
\section{Conclusion}
\label{sec:conclusion}

We present a novel approach for anonymizing speaker embeddings for privacy-preserving dementia detection. Our experiments show the potential of using prosody disentanglement to isolate specific attributes from speaker identity in low-resource settings. By training an embedding extraction model against prosodic extractors on an auxiliary dataset, we effectively minimise speaker identity information while preserving the diagnostic utility for dementia. We show the effectiveness of shuffling selected embedding dimensions for anonymization with an impact similar to classifier-dependent adversarial learning. Notably, our disentanglement approach outperforms the traditional adversarial training, showcasing the importance of the dataset size. Among our selected features, disfluency and articulatory features contribute to preserving dementia with little impact on the speaker's privacy. Our work is a first step towards privacy-preserving speaker embeddings for healthcare applications and we hope to inspire further refinements of these approaches.

\bibliographystyle{IEEEtran}
\bibliography{mybib}

\end{document}